\documentclass[superscriptaddress,aps,twocolumn]{revtex4}
\usepackage{amsfonts}
\usepackage{amsmath}
\usepackage{amssymb}
\usepackage{graphicx,color}
\usepackage{dcolumn}
\usepackage{times}
\usepackage{relsize}
\usepackage{mathrsfs}

\usepackage{svg}
\usepackage{float}

\DeclareSymbolFont{matha}{OML}{txmi}{m}{it}
\DeclareMathSymbol{\varv}{\mathord}{matha}{118}

\begin{document}
\title{Acoustic 
solitons 
in a periodic waveguide: theory and experiments
}

\author{I.~Ioannou Sougleridis}  
\affiliation{Laboratoire d’Acoustique de l’Universit\'{e} du Mans (LAUM), UMR 6613, Institut d'Acoustique - Graduate School (IA-GS), CNRS,  Le Mans Universit\'{e}, France}
\affiliation{Department of Physics, National and Kapodistrian University of Athens, Panepistimiopolis, Zografos, Athens 15784, Greece}
\author{O.~Richoux}  
\affiliation{Laboratoire d’Acoustique de l’Universit\'{e} du Mans (LAUM), UMR 6613, Institut d'Acoustique - Graduate School (IA-GS), CNRS,  Le Mans Universit\'{e}, France}
\author{V.~Achilleos}  
\affiliation{Laboratoire d’Acoustique de l’Universit\'{e} du Mans (LAUM), UMR 6613, Institut d'Acoustique - Graduate School (IA-GS), CNRS,  Le Mans Universit\'{e}, France}
\author{G.~Theocharis}  
\affiliation{Laboratoire d’Acoustique de l’Universit\'{e} du Mans (LAUM), UMR 6613, Institut d'Acoustique - Graduate School (IA-GS), CNRS,  Le Mans Universit\'{e}, France}
\author{C.~Desjouy}  
\affiliation{Laboratoire d’Acoustique de l’Universit\'{e} du Mans (LAUM), UMR 6613, Institut d'Acoustique - Graduate School (IA-GS), CNRS,  Le Mans Universit\'{e}, France}
\author{D. J.~Frantzeskakis}
\affiliation{Department of Physics, National and Kapodistrian University of Athens, Panepistimiopolis, Zografos,  Athens 15784, Greece}

\begin{abstract}
We study the propagation of high-amplitude sound waves, 
in the form of pulse-like solitary waves,
in an air-filled acoustic waveguide of periodically varying cross section.  
Our numerical simulations, solving the compressible Navier-Stokes equations in two dimensions,
 as well as our experimental results, strongly suggest that nonlinear losses, originated from vortex shedding (at the segment changes) are crucial in the dynamics of high amplitude pulses.
We find that, even in the presence of strong dissipation, the solitary wave roughly retains its characteristics (as described by the amplitude-velocity-width relations),  obtained by the derivation and analysis of an effective Boussinesq equation. 
In addition, we propose a transmission-line based numerical scheme, able to capture well the experimental results. The proposed design offers a new playground for the study of the combined effects of dispersion, nonlinearity and dissipation in air-borne acoustics, while, due to its simplicity, it can be extended to higher dimensions.
\end{abstract}
\maketitle

\section{Introduction}

Nonlinearity, dispersion and dissipation are well known to play a key role in wave propagation, and in many physical systems these mechanisms coexist. One of the most interesting phenomena which stems from the balance of dispersion and nonlinearity is the emergence of solitons, namely robust localized waves propagating undistorted in nonlinear dispersive media  \cite{ablowitz,solitons_1,solitons_otical,Dauxois,Remoissenet}. The existence, stability dynamics and interactions of 
solitons have been studied
extensively, both in theory and in experiments, in various physical contexts including 
water waves\,\cite{zabusky},
plasma physics\,\cite{plasma}, nonlinear optics\,\cite{kivshar}, atomic Bose-Einstein
condensates\,\cite{kevrekidis},
and so on.

Furthermore, solitons have also been theoretically predicted and experimentally observed in acoustics, both in solids and in fluids. 
Relevant studies have been performed in a variety of solid state settings,
such as mechanical lattices\,\cite{lattices,Pnevmatikos_3,JSV2}, granular chains\,\cite{granular2}, flexible elastic architected materials\,\cite{tournat} and crystalline solids\,\cite{crystalline}. On the other hand, acoustic solitons in fluids have been studied less extensively. 
One can argue that this is due to the fact that the natural nonlinearity in a fluid is weak, thus demanding relatively large systems to be excited, which, in turn, leads to 
the presence of unavoidable dissipation/losses. Also, generally, dispersion is very weak \cite{JSV1} in fluids and, only recently, 
various mechanisms (e.g., artificial inhomogeneities, resonances etc) have been used to induce strong dispersion effects.

Nevertheless, and despite the above mentioned drawbacks, there exist a number of works 
devoted to the study of generation and dynamics of solitons in air-filled acoustic waveguides. In fact, the pioneering work of Sugimoto {\it{et al.}}\,\cite{sugimoto_soliton_2,sugimoto_soliton_3,sugimoto_soliton_4,sugimoto_5,sugimoto_6} has 
paved the way for both theoretical and experimental studies of acoustic solitons 
in air. The main structure considered in these works is a waveguide sideloaded by a periodic array of Helmholtz resonators. In this case, the low frequency resonance induces  dispersion for the acoustic plane waves. The nonlinearity for high amplitude waves stems from the constitutive equations and the nonlinear dependence of the density on pressure. 
 
The propagating nonlinear pulses were found in an implicit form, and turned out to be close to Korteweg-de Vries (KdV) solitons \cite{ablowitz} in some asymptotic limit; both
numerical studies on the model proposed and experiments verified the theoretical predictions\,\cite{sugimoto_prl,sugimoto_review}.
The same structure was extensively revisited more recently,
with the use of 
an analytical model based on the transmission line approach, 
and numerical simulations. Both the analytical and numerical results 
were found to be in good agreement with pertinent experimental observations of 
solitons\,\cite{resonators,wave_motion,vassos_soliton}.
Other waveguide structures, 
loaded with arrays of elastic membranes 
or side holes, were also predicted 
to support solitons  
in airborne acoustics
\,\cite{kinezoula_dark,kinezoula_gap}. In these latter works, where the waveguide dispersion 
was introduced by periodicity, it was shown that the 
predicted solitons were in fact envelope solitons   
obeying effective nonlinear Schr\"{o}dinger (NLS)  equations.

In this work, our scope 
is to study the interplay between nonlinearity and dispersion, as well as the nature and role of dissipation, in a acoustic waveguide of periodically varying cross section.
In particular, not only we are 
interested in the generation and propagation of acoustic solitons in this setting, but we also wish to evaluate the influence of strong linear and nonlinear dissipation on the soliton  
characteristics, since all these 
phenomena occur generically
in airborne acoustics. Notice that here we use the term soliton in a loose sense, without implying complete integrability \cite{ablowitz}.

A brief description of our approaches and results, as well as the structure of our presentation, are as follows. First, in Section\,II, we describe our experimental setup and describe linear wave propagation in the periodic acoustic waveguide using the transfer matrix method (TMM), including the effect of viscous and thermal losses. To verify the uni-dimensional (1D) approximation we perform experiments using a low amplitude sinusoidal source. Then, we present experimental results for the formation of acoustic solitons in the periodic waveguide, using high amplitude square pulses, generated by a balloon explosion, as a source. The results are corroborated by numerically solving 
the two-dimensional Navier-Stokes equations. In Section\,III we consider an effective partial differential equation (PDE) that includes dispersive and nonlinear effects, thus ending up with 
a Boussinesq-type equation that describes weakly nonlinear and weakly dispersive waves. This Boussinesq model   
allows us to obtain exact analytical 
soliton solutions that can propagate in the periodic acoustic waveguide. Next, we use a "super-cell" transmission line (TL) approach to model the nonlinear wave propagation in the acoustic waveguide. This approach 
accurately captures the dispersive characteristics of the system, but also allows us to include both viscothermal losses and nonlinear attenuation due to acoustically induced vortex shedding at the locations where
cross sections change. Our numerical investigations  
confirm, at first the analytical predictions 
in the lossless regime, by verifying the soliton existence. We further demonstrate that for weak losses (linear and nonlinear) the soliton persists, despite the fact that its characteristics deviate slightly from the lossless theory. In Section\,IV, we study numerically and experimentally how the soliton characteristics, 
i.e., the relation between the amplitude, velocity and width, are affected by the weak and strong nonlinear dissipation.
Pertinent results are found to be in a good agreement with our experimental findings and  clearly demonstrate the generation and propagation of acoustic solitons,  
despite the presence of strong dissipation. 
Finally, in Section~V, we summarize our findings
and we present future research directions.


\section{Experimental observations using high amplitude pulses}

\begin{figure}[tbp]
\begin{center}
   \includegraphics[width=0.45\textwidth,]{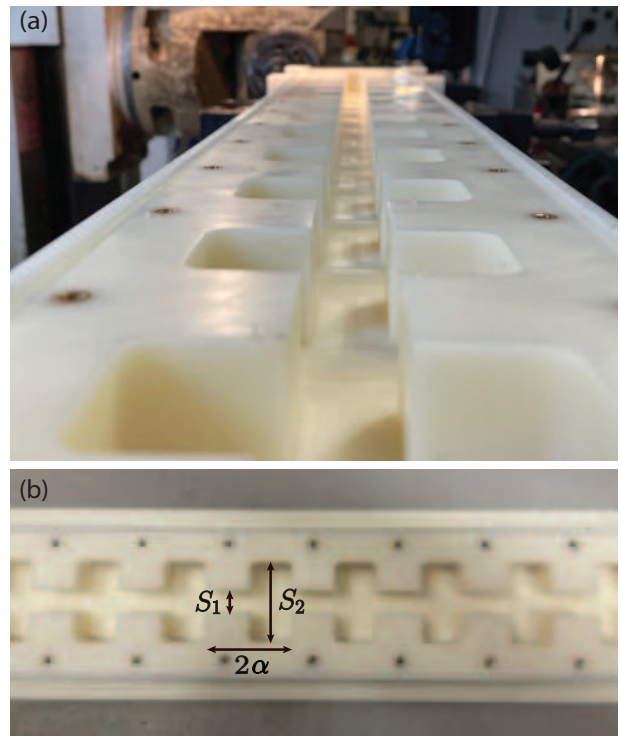}
  \caption{Experimental setup. (a) Side view of the acoustic waveguide. (b) Top view of the waveguide, where the two different cross sections $S_1$ and $S_2$ and the segment length are indicated. \label{physical setup}}
  \end{center}
\end{figure}

The acoustic system under consideration 
is composed of periodically arranged rectangular segments of length $\alpha=3.5$\,cm with two different widths $h_1=2$~cm and $h_2=7$\,cm. The two widths correspond to two different cross sections $S_1=6$\,cm$^2$, $S_2=21$\,cm$^2$, as shown in Fig.\,\ref{physical setup}.

In the experiments, the waveguide is closed on the top by a plexiglass plate, and microphones at different positions are used to measure the acoustic pressure field. 
The total waveguide structure can be as large as $N=34$ unit cells and it is closed at both ends. 
Acoustic waves are generated using two types of sources, namely a typical loudspeaker for the linear regime, as well as a balloon explosion to generate nonlinear pulses inside the waveguide. The explosion takes place in a cavity of length $15$~cm. Finally, we have used four different microphones PCB106B, carefully calibrated, placed at positions $x_1=0.0175$\,m, $x_2=0.35$\,m, $x_3=0.77$\,m and $x_4=1.33$\,m to measure the acoustic pulse  inside the periodic waveguide. 

\begin{figure}[tbp]
\begin{center}
   \includegraphics[width=0.4\textwidth]{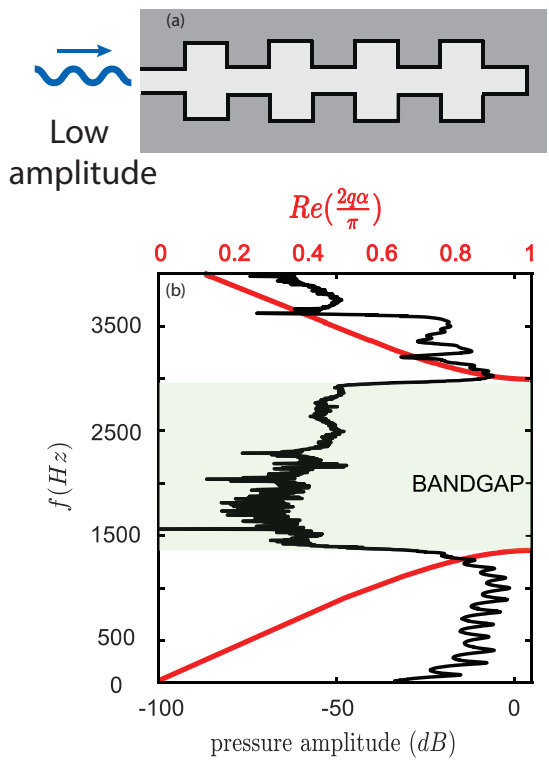}
  \caption{A sketch of a finite acoustic waveguide, with the low amplitude sinus source in panel (a). The TMM dispersion relation (\ref{transfer matrix}) (red line and red axis) along with the experimental pressure amplitude as a function of the frequency for a waveguide cavity composed of $N=14$ unit cells. 
  \label{experiment linear}}
  \end{center}
\end{figure}

We first perform experiments to quantify the dispersion properties of the system in the linear regime. In particular we excite the system with $N=14$ unit cells, using monochromatic waves and we measure the spectrum at the end of the system illustrated in Fig.\,\ref{experiment linear}(a). The corresponding results are shown in Fig.\,\ref{experiment linear}(b) with the solid (black) line . Here we see that at low frequencies
we identify the modes of the structure within the propagating band and a clear band gap is detected (shaded green).
In fact, within the first branch we measure 13 peaks in the experimental data (black line) corresponding to the resonant frequencies of the cavity. 
According to the theory, for sufficiently low frequencies, such that only the fundamental acoustic mode is propagating, waves are described by the one-dimensional (1D) Helmholtz equation and the predicted dispersion relation is 
(see Appendix\,\ref{appendix}):
\begin{align}
    &\cos (2 q \alpha)=\cos^2 \left(k\tilde{ \alpha}\right)
    -\frac{S_{1}^{2}+S_{2}^{2}}{2S_{1} S_{2}} \sin^2 \left(k \tilde{\alpha}\right).\label{transfer matrix}
\end{align}
In the latter expression  $k\tilde{\alpha}$ is the dimensionless wavenumber, which includes the added length due to the abrupt change $\tilde{\alpha}=\alpha+\delta\alpha$ \cite{correction_1,correction_2,correction_3}. $2q\alpha/\pi$ denotes the dimensionless Bloch wavenumber.

\begin{widetext}
\begin{figure*}[tbp]
\begin{center}
   \includegraphics[width=1\textwidth]{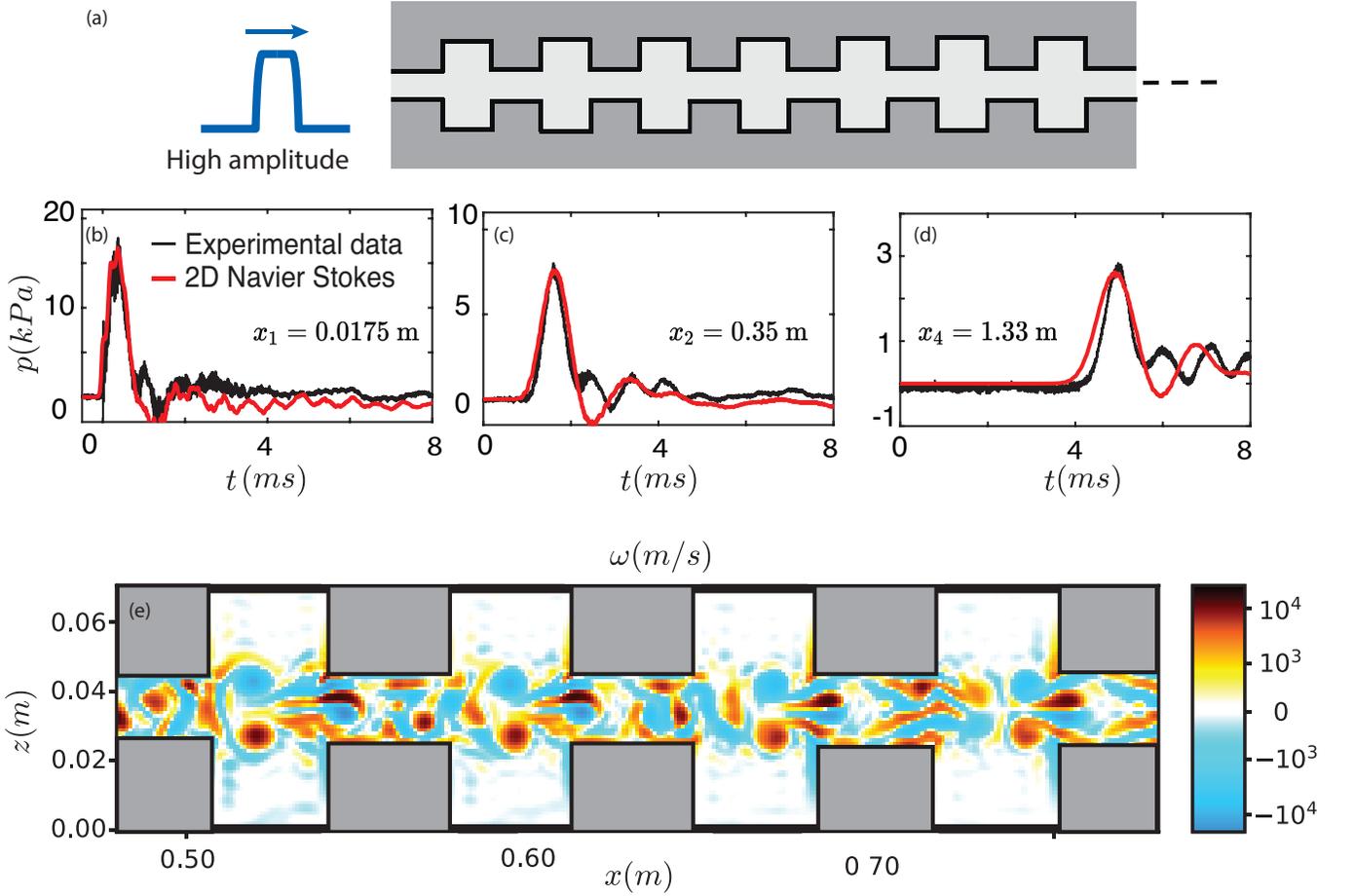}
  \caption{(a) A sketch of a semi-infite periodic waveguide with a high amplitude square pulse. Presented in panels (b)-(d) are the temporal profiles of the pressure derived from experiments (black line) and by solving the NS equations (solid red line) at positions $x_{1,2,4}$ respectively. Finally, panel (e) illustrates a zoom of the vorticity field after the arrival of the pulse around $x=0.6$\,m.
  \label{experiment nonlinear}}
  \end{center}
\end{figure*}
\end{widetext}

We now proceed to experiments with high amplitude pulses in a lattice with $N=34$ unit cells. Such an acoustic source is generated experimentally by a balloon explosion at one end (inside) the waveguide as illustrated in Fig.\,\ref{experiment nonlinear} (a). The corresponding generated time signal of the pressure has the form of a square pulse as it shown in panel (b) of Fig.\,\ref{experiment nonlinear}. We then measure the pressure at three different positions $x_1=0.0175$\,m, $x_2=0.35$\,m, $x_3=0.77$\,m and $x_4=1.33$\,m.
Typicals example of the experimentally obtained temporal profiles are shown in panels (c)-(d) of Fig.\,\ref{experiment nonlinear}, and are depicted  with the (black) thick line.


The initial  explosion of the  balloon has generated a pulse which propagates in the waveguide. Since the system is dispersive, without nonlinearity the pulse should substantially be broadened. However here we observe that it propagates without substantial modification of its shape, clearly indicating a nonlinear wave. In addition a tail of smaller amplitude waves is preceding it. Although the shape of the pulse appears unchanged, its amplitude decreases. It is important to note that the high amplitude pulse shows a much faster decay than the tails. This suggests that amplitude dependent losses, thus nonlinear, are present in the system on top of the viscothermal (linear) attenuation.
To further investigate this observation, we perform numerical simulations in the
framework of the full 2D compressible Navier-Stokes (NS) equations.

We use a large amplitude pulse at the left boundary of the waveguide and the corresponding pressure signals obtained by the simulation are superimposed together with the experimental result in panels  (b)-(d) of  Fig.\,\ref{experiment nonlinear}  with the red (dashed) line. 
The agreement between simulation and experiment ensures that the physical phenomena are well captured by the numerical model and reveal the nonlinear nature of the wave propagation. In addition the physical mechanism of nonlinear losses is understood by monitoring the vorticity field, shown in Fig.\,\ref{experiment nonlinear} (e), when the pulse passes from the region of the waveguide around $x=0.6$\,m. 
 It is clear that vortices are generated close to the change of cross section and we have also verified that these vortices move. It is known that vortex generation and shedding induces nonlinear loss of acoustic energy: one can imagine that a part of the acoustic energy is transformed into vortical motion responsible for the strong decay of the nonlinear pulse.

Our experimental results show that large amplitude pulses in the proposed periodic waveguide are prone to three basic phenomena: dispersion, nonlinearity and dissipation (linear and nonlinear). 
In the next section, an effective 1D PDE is constructed which admits solitary wave solutions whose characteristics (amplitude, width and velocity) match the ones of the experimentally observed nonlinear pulses.
We also propose a simplified 1D model which captures all the aforementioned phenomena and gives a nice agreement with experimental results. 

\section{1D modeling}

\subsection{Effective Boussinesq equation and soliton solutions}
Here, we wish to  formulate an effective 1D PDE which captures both phenomena of dispersion and 
nonlinearity, allowing the study of acoustic 
solitons in the periodic waveguide. 
To incorporate the dispersion induced by the change of cross sections we follow a procedure similar to Refs.~[28, 33]. Knowing the analytical form of the dispersion relation given by Eq.\,(\ref{transfer matrix}), we construct the corresponding linear operators and their coefficients which reproduce the correct dispersion. The detailed calculation is found in Appendix\,\ref{app2}.
Regarding nonlinearity, it is well known that high amplitude acoustic waves feature nonlinear effects. To first order, the nonlinearity of propagating waves in the bulk is due to the nonlinear relation between density and pressure through the equation of state in a quiescent fluid. In one dimension, one of the commonly used equations which captures this effect for acoustic waves is the 1D Westervelt equation, whose nonlinear term comes from an amplitude dependent correction of the sound celerity \cite{hamilton}. 

By combining this correction along with the aforementioned dispersive terms we propose the following Boussinesq type equation:
\begin{equation}
    p_{tt}-\tilde{c}^2 p_{xx}-\beta_m p_{xxtt}
    -\beta_x p_{xxxx} -b (p^2)_{{tt}}=0
\label{nonlinear homogenization}.
\end{equation}
Here, subscripts denote partial derivatives with respect to time $t$ and propagation distance $x$,  $\tilde{c}=\tilde{c}_0/\tilde{\alpha}$ is the effective speed of sound, while $\beta_x$ and $\beta_m$ are the dispersion coefficients. The nonlinearity coefficient is given by $b=\beta/\varrho_{0} c_{0}^{4}$,  
with $\varrho_{0}$ being the air density evaluated at the equilibrium state and $\beta=1.2$ for air. Our approach assumes that the correction of the sound speed does not depend on the width of the waveguide. In Eq.\,(\ref{nonlinear homogenization})
the linear coefficients $\tilde{c},\,\beta_m$ and $\beta_x$ are found following the procedure as in Refs.\,\cite{cornaggia,maurel} (see also Appendix).

Notice that Boussinesq type equations belong to the 
well studied class of weakly dispersive and weakly nonlinear PDEs \,\cite{ablowitz}; 
as such, they have been used  
in studies of waves in various contexts, such as  
shallow water waves \cite{ablowitz,solitons_1,Dauxois},
mechanical lattices and electrical transmission lines \cite{Remoissenet,Dauxois,Kofane}, acoustics~\cite{vassos_soliton}, and so on.

We now proceed with the derivation of soliton solutions of Eq.\,(\ref{nonlinear homogenization}), which are possible 
in the regime where the dispersion and
nonlinearity terms are of the same order. 
In this regime, travelling wave solutions of the above equation can readily be obtained by introducing the ansatz $p(x,t) =P(\xi)$, where $\xi=x-\mathrm{v} t$ (with $\mathrm{v}$ being the velocity of the wave). Introducing this ansatz in 
Eq.\,(\ref{nonlinear homogenization}), and assuming vanishing boundary conditions for $P(\xi)$, namely $P \to 0$ as $|\xi| \to \infty$, we obtain the following ordinary differential equation (ODE)  for $P(\xi)$:
\begin{equation}
     P^{\prime\prime}+\mathcal{B} P-\mathcal{D} P^2=0, \label{ODE}
\end{equation}
where primes 
derivatives with respect to $\xi$, while the parameters involved in the above ODE are given by  $$\mathcal{D}=-(b\mathrm{v}^2)/(\beta_x+\beta_m \mathrm{v}^2), 
\quad 
\mathcal{B}=(\mathrm{v}^2-\tilde{c}^2)/(\beta_x+\beta_m \mathrm{v}^2)
$$ 
depend on both the dispersive and nonlinear coefficients. A 
straightforward analysis\,\cite{vassos_soliton} shows that 
the homoclinic solution of the ordinary differential equation \,(\ref{ODE}), which exists for $\mathrm{v}>\bar{c}$, corresponds to the following soliton solution of Eq.\,(\ref{nonlinear homogenization}):

\begin{equation}
  p(x, t)=A\operatorname{sech}^{2}\left(w\left(x-\mathrm{v} t\right)\right),\label{boussinesq soliton}
\end{equation}
where the amplitude $A$ and the inverse width $w$ of the soliton are given as follows
\begin{equation}
A=\frac{3\left(\mathrm{v}^{2}-\tilde{c}^{2}\right)}{2 b \mathrm{v}^{2}},\quad
w=\frac{1}{2}\left(\frac{\mathrm{v}^{2}-\tilde{c}^{2}}{\beta_x+\beta_m \mathrm{v}^2}\right)^{1/2}.\label{solamp}
\end{equation}
Notice that the above solution is characterized by one free parameter, the velocity $\mathrm{v}$. 
Below, we propose a 1D numerical scheme which verifies the propagation of solitary waves given by Eq.\,(\ref{boussinesq soliton}) in the periodic waveguide and which also captures experimental observations when losses are introduced.

\subsection{1D Supercell  model}
Our modeling relies on the transmission line (TL) theory which is widely used in acoustics based on the electroacoustic analogy. This analogy has also been successfully used to describe nonlinear propagation in acoustic waveguides \cite{ vassos_soliton,kinezoula_dark,kinezoula_gap}.

\begin{figure}[tbp]
\begin{center}
   \includegraphics[width=0.40\textwidth]{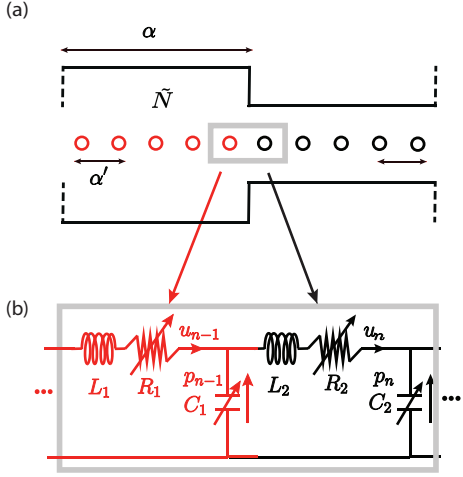}
    \vspace{-1.25 cm}
  \caption{(a) A unit cell of the periodic waveguide and its supercell TL representation. Depicted in red is the last point before the change of cross section for the small waveguide segment, while depicted in black is the first point of the big waveguide segment. (b) The gray box shows the supercell TL and its discrete points in the region change of cross section of the periodic waveguide. \label{supercell representation}}
  \end{center}
\end{figure}

Assuming long waves (where the 1D approximation holds) we discretize each segment of the periodic waveguide of length $\alpha$ in a total number of $\tilde{N}$ points, as presented in Fig.\,\ref{supercell representation}(a). 
According to the electroacoustic analogue, 
the propagation along a segment is modeled by a circuit for the "voltage" $p_n$ (corresponding to the pressure) and the "current" $u_n$ 
(corresponding to the acoustic flux as illustrated in Fig.\,\ref{supercell representation}(b)). To take into account nonlinear effects induced by high amplitude acoustic signals, 
the inductance and the capacitance at the $n$-th point 
assume the following form \cite{vassos_soliton,kinezoula_dark,kinezoula_gap}:
\begin{equation}
L_{n}=\varrho_{0} \alpha^\prime / S_n, \quad C_{n}\approx\frac{S_n \alpha^\prime}{\varrho_{0}c_{0}^{2}} \left(1-2\frac{\beta}{\varrho_{0} c_{0}^{2}}p_n\right)
\end{equation}
where $S_n$ is the cross section of the waveguide at the $n$-th point, and $\alpha^\prime=\alpha/\tilde{N}$ is the distance between two points (see Fig.\,\ref{supercell representation}(a)). We then use Kirchhoff’s voltage and current laws for two successive lattice points, and derive the following set 
of differential-difference equations: 
\begin{align}
& L_{n}\frac{du_{n}}{dt}+R_{n}u_{n}=p_{n-1}-p_{n},\label{voltage1}\\
&L_{n+1} \frac{du_{n+1}}{dt}+R_{n+1}u_{n+1}=p_{n}-p_{n+1},\label{voltage2}\\
&\frac{d}{dt} (C_{n}p_{n})=u_{n}-u_{n+1},\label{current1}
\end{align}
where $R_n$ quantifies the total losses of the system.
To choose the number of points $\tilde{N}$ within each segment, we compare the dispersion relation  characterizing  the transmission line in the linear regime with the TMM (Eq.\ref{Bloch dispersion}) and here we find that $\tilde{N}=15$ is sufficient (see Fig.\,\ref{dispersion2}(a) in Appendix\,\ref{app2}) to accurately capture the dispersion of the periodic waveguide.

The resistance used in Eqs.\,(\ref{voltage1}-\ref{voltage2}) has the following nonlinear form \cite{vassos_nonlinear,hirschberg}
\begin{align}
R_n=R_{0n}+g|u_{n}|. \label{nonlinear losses}
\end{align}
Here, $R_{0n}$ represent the linear viscothermal losses occurring due to the friction at the boundaries, which depend on the cross section of the waveguide. This linear loss term  is determined from Eqs.\,(\ref{wavenumberlosses}-\ref{impedancelosses}). 
As we have shown in the previous section, vortex shedding leads to nonlinear dissipation of acoustic energy, quantified by the nonlinear coefficient $g$ in Eq.\,(\ref{nonlinear losses}) 
\cite{sugimoto_review,wave_motion,vassos_nonlinear,hirschberg,kinezoula_nonlinear_holes}.
Although some explicit expressions for the parameter $g$ exist in the literature (see, e.g., Ref.\,\cite{hirschberg}), here 
we optimise to fit this coefficient according to our experiments. 

\begin{figure}[tbp]
\begin{center}
   \includegraphics[width=0.45\textwidth]{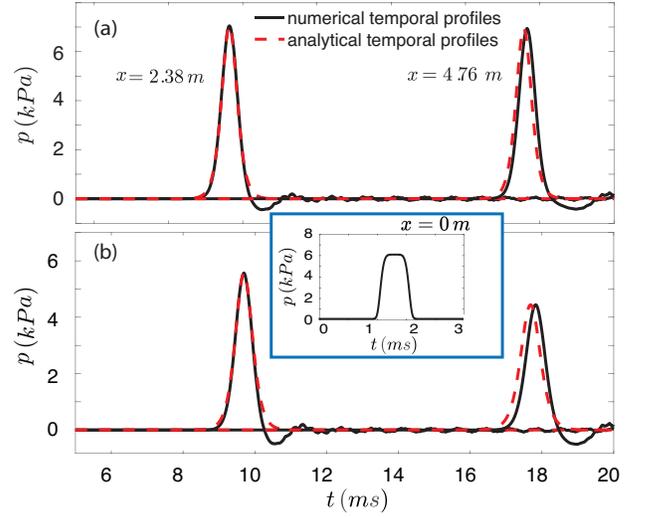}
  \caption{Evolution of a square pulse in the acoustic waveguide, discarding dissipation (a) and considering viscothermal losses and weak nonliner losses (b). Illustrated in both panels are the temporal profiles (black line) of the solitary waves at positions $x=2.38\,m$ and $x=4.76\,m$. The dotted red lines corresponds to a fit of the effective soliton solution, Eq.~(\ref{boussinesq soliton}), for $x=2.38\,m$, while the second one is shifted in time with the use of the soliton velocity.
Parameter values correspond to the experimental ones for a cross section ratio $S_2/S_1=7/2$.  
\label{supercell_soliton}}
  \end{center}
\end{figure}


To confirm soliton propagation in the waveguide, we integrate Eqs.\,(\ref{voltage1}-\ref{current1}) using 
the (time dependent) boundary condition at $x=0$\,m given by the following equation 
\begin{equation}
   p(t)= P_0\exp \left[-\left(t /w_0\right)^{6}\right]. \label{super gaussian}
\end{equation}
as shown in the inset of Fig.\,\ref{supercell_soliton}. We use the parameter values  $P_0=6$\,kPa and $w_0/2\approx500$\,Hz.

The evolution of such a boundary condition along the waveguide (i.e., along $x$) for the lossless case ($R_n=0$), features some transient phenomena followed by the propagation of a solitary wave. The pressure at a distance  $x=2.38$\,m is shown in Fig.\,\ref{supercell_soliton}(a), with the solid (black) line and has the form of a localised pulse followed by a small amplitude tail.
The analytical form of the soliton solution of Eq.\,(\ref{boussinesq soliton}) is superimposed in Fig.\,\ref{supercell_soliton}(a) with dashed lines. We use the maximum of the numerical solution as the value of the amplitude $A$, which fixes the width and the velocity through Eq.\,(\ref{boussinesq soliton}). 
 In addition, we plot the predicted analytical soliton at $x=4.76$\,m using a time delay given by its theoretical velocity. The analytical solution is in excellent agreement with the pertinent numerical result obtained in the framework of Eqs.\,(\ref{voltage1}-\ref{current1}).

To further investigate the role of dissipation, we use the same boundary conditions as in Fig.\,\ref{supercell_soliton}(a) but now using $R_n \ne 0$ with $g=b\times1.8\, 10^{-3}$. Note that this value of $g$ corresponds to a slower amplitude decay than our experimental and numerical observations in Section II. Hence, from now on we will refer to such small values of $g$ as the weakly dissipative regime.

Even in the presence of losses, the soliton solution remain a good approximation of the nonlinear pulses. The relevant results are shown in Fig.\,\ref{supercell_soliton}(b) for the same positions as in panel (a). We still observe the formation and propagation of a solitary wave inside the periodic waveguide although its amplitude decreases during propagation. 
Henceforward, we assume that the solitons persist under the influence of dissipative effects but with modified characteristics (amplitude-width/velocity). To justify this assumption, we compare each numerical pulse with the corresponding analytical soliton obtained using the two different numerical amplitudes. In this case we use a time delay given by the average of the velocities of each theoretical solution. 
At $x=2.38$\,m the numerical pulse is well captured by the corresponding lossless soliton solution as at $x=4.76$\,m, with discrepancy in the velocity.
This comparison confirms the generation and propagation of acoustic solitons, described by Eq.\,(\ref{boussinesq soliton}) under the influence of weakly dissipative phenomena, but with slightly altered characteristics.

It is worth to note that in both simulations the soliton is followed by an oscillatory tail. This tail could be a remnant of the effect of the boundary condition, since it does not have the exact shape of a soliton. Thus, as the initial acoustic pulse-like wave propagates along the waveguide, it reorganises itself into a soliton and, during this process, an additional radiation is emitted. However, we have found that in the lossless case, the two are well separated after sufficient propagation distance, while in the dissipative case the tail remains attached to the solitary wave.

\section{Experimental solitons}

\begin{figure}[tbp]
\includegraphics[width=0.45\textwidth]{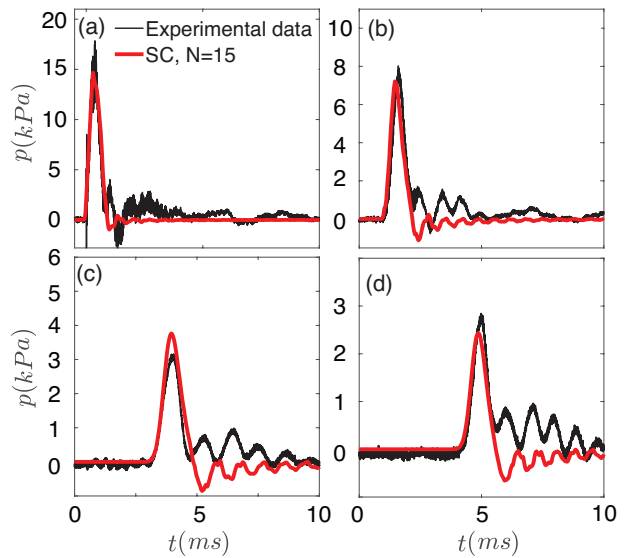}
 \caption{Comparison of the experimental pressure time signals and the numerical super-cell transmission line (solid red) at the positions of the microphones $x_{1,2,3,4}$  at panels (a), (b), (c) and (d) respectively. \label{comparison}}
\end{figure}

In this Section, we focus on the characterization and dynamics of the experimentally observed nonlinear pulses. In Section III we have demonstrated the propagation and formation of solitons in the acoustic waveguide despite the presence of \textit{weak} dissipative effects. Our goal is to assess if the experimental nonlinear pulses are solitons propagating under the influence \textit{strong} losses and if so, to compare their characteristics (amplitude-velocity/width) with the lossless theory and the weakly dissipative solitons.  We first need to identify the relevant value of $g$ corresponding to our experimental findings.  
To do so, we repeat the numerical simulations of Section III and find the value of $g=b\times3.8\, 10^{-2}$ that captures best the amplitude decay of the experimental pulse. 
We illustrate the corresponding numerical results with the experimental temporal profiles at all the measurement positions, $x_{1,2,3,4}$ in (\ref{comparison}) (a)-(d). In fact, the obtained value is not far away from the typical values given in the literature \cite{hirschberg,vassos_nonlinear,kinezoula_nonlinear_holes}. Here we would like to stress that the proposed simplified supercell model, other than a slight difference in amplitude captures the experimentally measured pressure profile quite well, especially as concerns the core of the pulses.

As we have seen in Sections II and III both experimental and numerical nonlinear pulses are followed by an oscillatory wave. Note that in Fig.\,\ref{comparison} there is a clear phase difference between the experimental and numerical tails following the pulse. The most plausible explanation lies in the generation of the experimental source. In the experiments, an additional cavity is used for the balloon explosion, which is not included in the numerical simulations.
The presence of this cavity may be responsible for the observed phase difference between the tails.

\begin{figure}[tbp]
\begin{center}
  \includegraphics[width=0.4\textwidth]{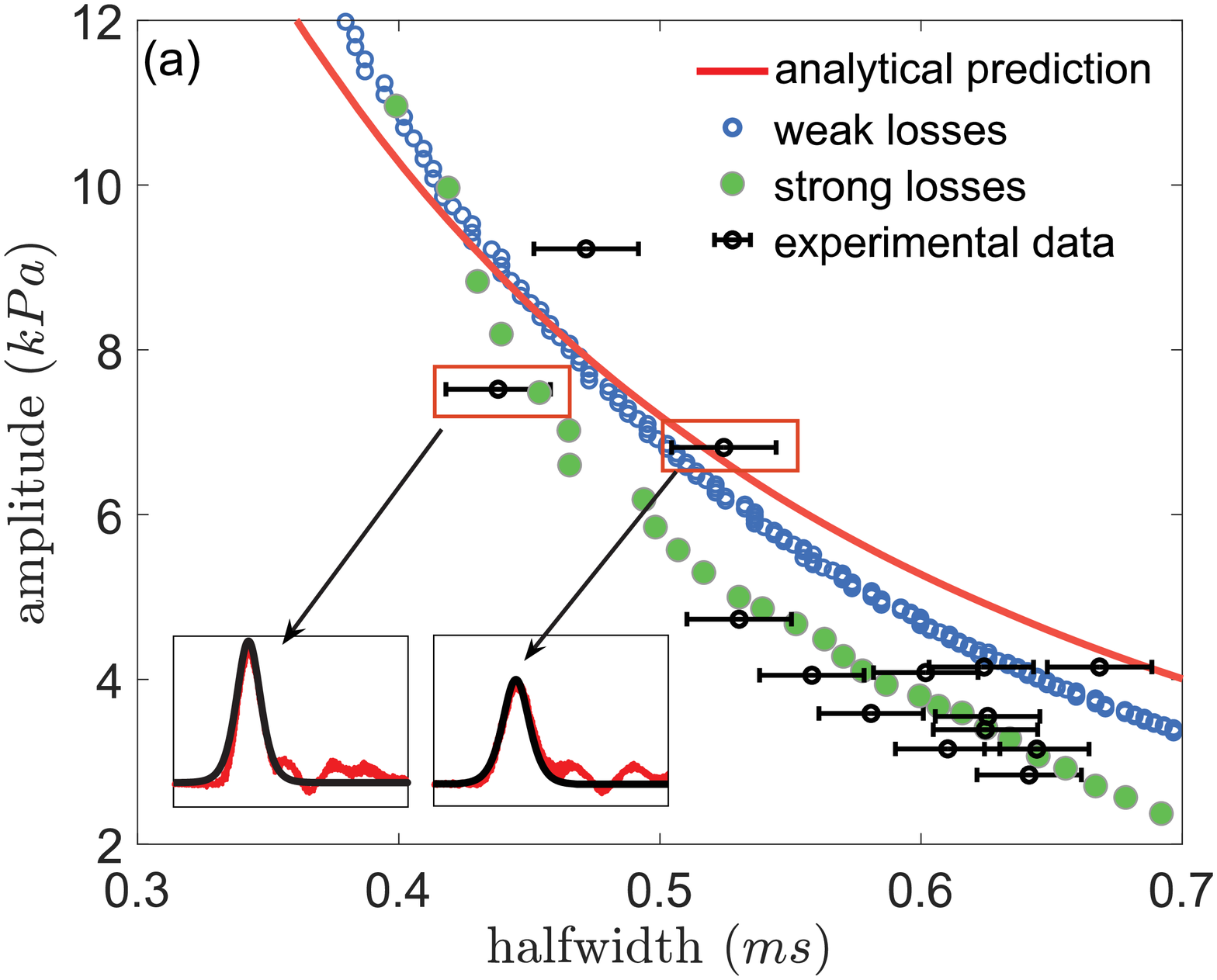}
  \includegraphics[width=0.4\textwidth]{
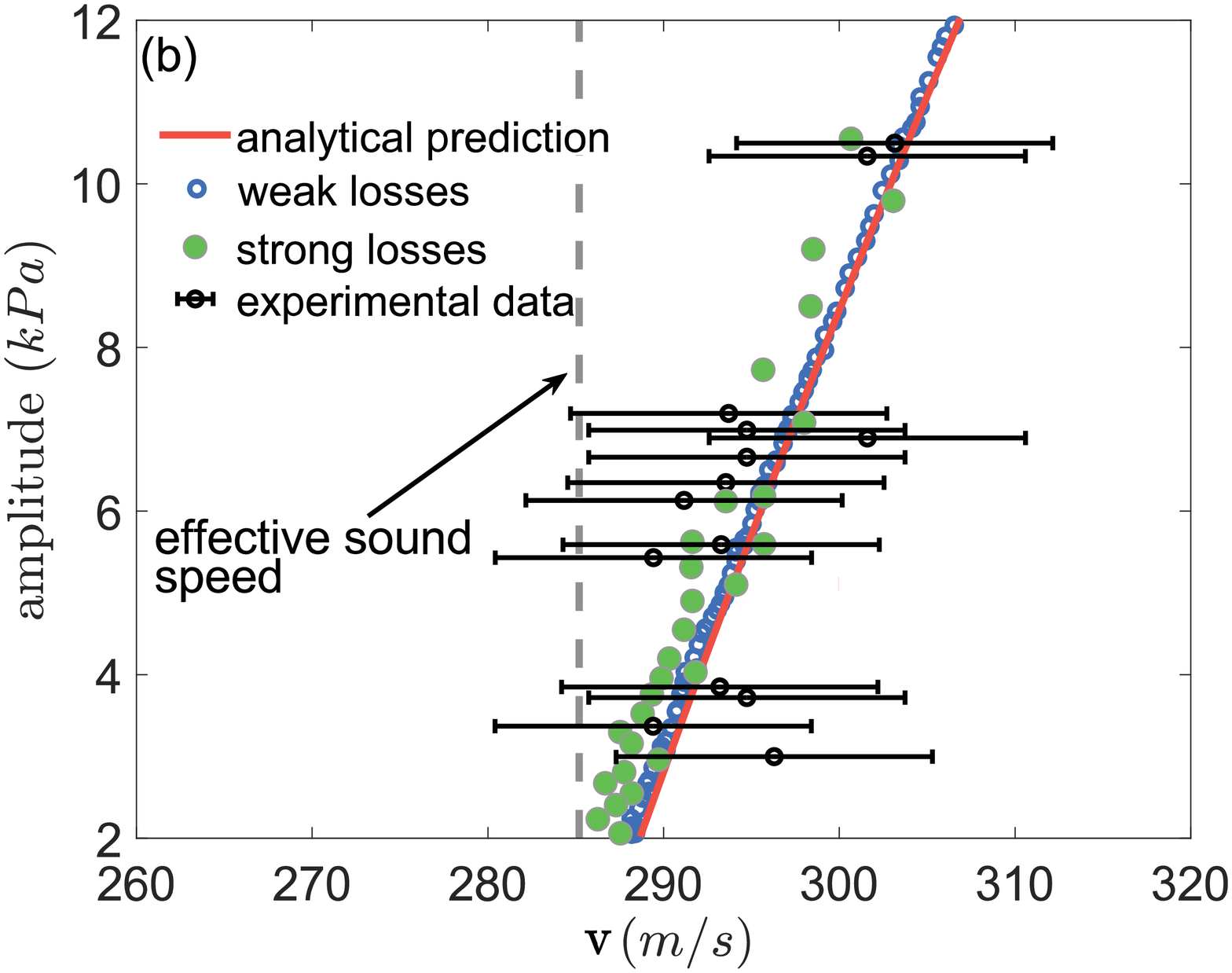}
  \caption{ (a) The soliton amplitude as a function of the soliton halfwidth
  as predicted by the analytical expression (\ref{boussinesq soliton}) (red), the supercell simulations, with weak (blue) or strong (green circles) nonlinear losses, and the experimental data (black). Depicted in the insets are the soliton's  
  temporal profiles as observed in the experiments (red) and 
  predicted by expression (\ref{boussinesq soliton})
  (black).  (b) Similar to panel (a) but now for the soliton amplitude as a function of the soliton velocity. Depicted in gray is the linear effective speed of sound.
  \label{halfwidth}}
   \end{center}
\end{figure}

Performing multiple experiments with different sources (balloon explosions) we have measured pulses of different initial shapes in the periodic acoustic waveguide. This allows us to further investigate the  signature of the solitons, as given by its amplitude-velocity/width dependence.
We first consider the relation between the amplitude and halfwidth of the nonlinear pulse. The results are shown in Fig.\,\ref{halfwidth}(a) where
the experimentally measured pulse amplitude/width relation, is shown in black dots.
The experimental results follow the same tendency as the theoretical predictions (red line). In addition as it is observed in the inset of Fig.\,\ref{halfwidth}\,(a), the measured pulse 
core fits very well with the corresponding analytical soliton solution. 
The small discrepancy between (lossless) theory and experiments indicates that nonlinear losses result in more localised (thinner) pulses. Such an effect has also been observed in earlier works by Sugimoto et al \cite{sugimoto_review}. 

In order to further investigate the influence of losses, and thus the difference between experiments and analytics, we corroborate the amplitude-halfwidth relation with numerical simulations.
In the case of \textit{strong} losses (green circles) the numerical results quite accurately capture our experimental findings, certifying the proposed supercell TL scheme.
On the other hand, the curve corresponding to \textit{weak} losses (blue circles) appears to be closer to the analytical one.
In this case, dissipative effects can be considered as a small perturbation \cite{kima,dissipative_2,dissipative_3,dissipative_4} which affects the soliton parameters, that become slowly-varying functions of time. 
In fact, the strength of the dissipative effects can be characterized by the amplitude decay per unit cell. 
As long as this quantity is small, the evolution can be considered as \textit{adiabatic} \cite{kima} and, in this case, the relation of the soliton's amplitude and halfwidth, is weakly modified from
the analytical solution of Eq.\,(\ref{boussinesq soliton}).

The second important relation which characterizes solitons of the Boussinesq type is the dependence of the velocity on the amplitude, indicating the solitons supersonic nature. We present the velocities obtained experimentally as a function of  amplitude in
 Fig.\,\ref{halfwidth}(b) in circles (black) (measurement details are given n the Appendix). It is thus readily observed that the experimental pulses are supersonic, since the vertical dashed line in Fig.\ref{halfwidth}(b) 
corresponds to the linear speed of sound $\tilde{c}$.
More importantly, the experimental velocities remain 
close to the analytical prediction, which is confined in the error bar area. These results strongly suggest that the 
acoustic pulse-shaped waves observed in the experiments 
can safely be characterized as solitons, in accordance to our theoretical predictions.
 For completeness, we complement the above experimental results with numerical simulations (see Appendix for details on the calculation method). As in Fig.\,\ref{halfwidth}(a) weak losses (blue circles) appear to have a minor impact on the velocity of the soliton, verifying our assumption of adiabatic changes on the soliton characteristics. Furthermore, even in the presence of strong nonlinear losses the velocity follows very well the analytical lossless prediction and the proximity of the numerical results with experiments further justifies our transmission line model of Eqs.\,(\ref{voltage1})-(\ref{current1}).

\section{Conclusion}

We have studied high amplitude wave propagation in an air-filled acoustic waveguide with a periodic change of cross section. 
First, we have confirmed, both numerically and experimentally, that for sufficiently low frequencies our experimental setting 
features an effectively one-dimensional behavior.
Then, we have derived 
an effective wave equation to capture both dispersive and nonlinear effects. This equation, which was found to be of the Boussinesq type, allowed us to obtain exact  
analytical soliton solutions. 
Our theoretical prediction concerning the existence of acoustic solitons in our setting was corroborated by  
numerical simulations and experimental results. 
In particular, in the simulations, we considered both an effective supercell transmission line model, as well as the two-dimensional Navier-Stokes equations. We have also studied the effect of losses, and characterized their effect systematically, using both theory and experiments. In our experiments, we investigated thoroughly the soliton characteristics (velocity and width) and the obtained results were found to be in good agreement with corresponding ones obtained analytically and numerically.

Our methodology and results suggest a number of interesting themes for future investigations. These include the study of solitons and other types of nonlinear waves both in quasi one-dimensional waveguide structures, e.g., in relevant nonlinear acoustic metamaterials, and    
in waveguide networks of higher dimensions.

\appendix
\section{Transfer Matrix Method}
\label{appendix}
To study the linear properties, and in particular to identify the dispersion relation of the periodic waveguide, we use three different methods: 
(a) the analytical expression derived from the transfer matrix method (TMM), (b) numerical simulations based on finite elements method (FEM), and (c) experiments, by measuring the frequency response of our experimental setup.
Reagarding the TMM, the starting point of our analysis relies on the consideration of an ideal fluid, neglecting viscosity and other dissipative
terms. 
Moreover, we are interested in the regime of 
sufficiently low frequencies, such that only the fundamental acoustic mode of the waveguide is propagating. In this regime, the acoustic wave propagation is 
described by the 1D Helmholtz equation for the pressure \cite{kinsler,morse}:
 \begin{equation}
     \frac{d^2\hat{p}(x)}{d x^2}+k^2\hat{p}(x)=0 , \label{Helmholtz}
 \end{equation}
where $k=\omega/c_0$ is the wavenumber, while $\omega$ is the frequency and $c_0$ the speed of sound in air.  

\begin{figure}[h!]
 \vspace{1 cm}
\begin{center}
   \includegraphics[width=0.45\textwidth]{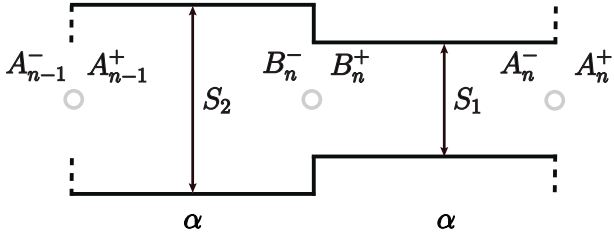}
  \caption{A unit cell of the periodic acoustic waveguide composed of cross sections $S_1$ and $S_2$. \label{unit_cell}}
  \end{center}
  \vspace{-0.5 cm}
\end{figure}

Equation~($\ref{Helmholtz}$) leads to the following plane wave solution, 
\begin{equation}
    \hat{p}(x)=A\mathrm{e}^{\mathrm{i}k x}+B\mathrm{e}^{-\mathrm{i}k x},
\end{equation}
while, accordingly, the flux velocity is given by:
\begin{equation}
    \hat{u}(x)=\frac{S_j}{\rho_0 c_0}\left[B\mathrm{e}^{-\mathrm{i}k x}-A\mathrm{e}^{\mathrm{i}k x}\right],
\end{equation}
where $\rho_0$ is the density of air, $S_j=w h_j$ is the cross section of the waveguide considered here to be rectangular, 
with $w$ and $h_j$ denoting its 
height and width 
respectively and $j=1,2$  denotes waveguide 
segments of cross sections $S_1$ and $S_2$ respectively.
Now, let us consider the unit cell depicted in Fig.\,\ref{unit_cell},
composed of a rectangular waveguide segment of length $\alpha$ and cross section $S_1$, connected with 
another waveguide segment of length $\alpha$ and cross section $S_2$. The pressure and the flux velocity between points $A^-_n$ and $B^+_n$ and between points $B^-_n$ and $A^-_{n-1}$ inside the waveguide, separated by a distance $\alpha$  are connected with  
the following transfer matrix:
\begin{equation}
\mathbf{T}_j=
\left[\begin{array}{ll}
\cos \left(k\tilde{\alpha}\right)&\mathrm{i}Z_j\sin\left( k \tilde{\alpha}\right)\\
\frac{\mathrm{i}}{Z_j}\sin \left(k\tilde{\alpha}\right)&\,\,\,\,\,\cos \left(k \tilde{\alpha} \right)
\end{array}\right],
\end{equation}
where $Z_j=\rho_0 c_0/ S_j$ is the waveguide impedance and $k\tilde{\alpha}$ is the dimensionless wavenumber by taking into account the added length \cite{correction_1,correction_2,correction_3}.
Thus, the pressure and the flux velocity between  $A^-_n$ and $A^+_{n-1}$ inside the waveguide, separated by a distance $2\alpha$, are connected 
with the following the transfer matrix
\begin{widetext}
\begin{align}
\left[\begin{array}{ll}
 p(x^{A}_{n-1})\\
 u(x^{A}_{n-1})
\end{array}\right]=\mathbf{T}\left[\begin{array}{ll}
 p(x^{A}_{n})\\
 u(x^{A}_{n})
\end{array}\right],
\quad 
\mathbf{T}=
\left[\begin{array}{ll}
 \cos^2{(k\alpha)}-\frac{Z_2}{Z_1}\sin^2{(k\tilde{\alpha})}&\mathrm{i}(Z_1+Z_2)\sin{(k\tilde{\alpha})}\cos{(k\tilde{\alpha})}
 \\
       ~~~\mathrm{i}\frac{Z_1+Z_2}{Z_1 Z_2}\sin{(k\tilde{\alpha})}\cos{(k\tilde{\alpha})}&~~~\cos^2{(k\tilde{\alpha})}-\frac{Z_1}{Z_2}\sin^2{(k\tilde{\alpha})}
\end{array}\right]
.
\end{align}
\end{widetext}


Now using the result for the transfer matrix $\mathbf{T}$ we can readily write a relation between the pressure and the velocity between two consecutive unit cells which reads
\begin{equation}
    \mathbf{{X_{n-1}}}=\mathbf{T} \mathbf{X_{n}}, \label{transfer matrix def}
\end{equation}
where $\mathbf{{X_{n}}}=\begin{bmatrix}
p(x^A_{n})& u(x^{A}_{n})
\end{bmatrix}^T$, is the Bloch eigenvector for the waveguide segment with cross section $S_A$ and for the $n$ unit cell of the periodic waveguide.
Assuming an infinite periodic waveguide with translation invariance, Eq. ($\ref{transfer matrix def}$) must hold for any $n$. Thus we may solve this equation by using a Bloch like anstaz \cite{soukoulis}. According to this ansatz the state vector will be given by
\begin{equation}
    \mathbf{X_{n}}=\mathbf{A}\mathrm{e}^{-\mathrm{i}2 q n \alpha},
\end{equation}
 where $\mathbf{A}$ is the column eigenvector.
Substituting this expression into equation ($\ref{transfer matrix def}$) we arrive to the following eigenvalue equation for the unknown phase factor $q\alpha$
 \begin{equation}
   \mathbf{T}\mathbf{A}=\mathrm{e}^{\mathrm{i}2q \alpha}\mathbf{I}\mathbf{A}. \label{eigevalue eq}
\end{equation}
This means that $\mathrm{e}^{\mathrm{i}2q \alpha}$ is an eigenvalue of the transfer matrix $T$. One of the properties of the transfer matrix, due to time reversal symmetry, is that its determinant is equal to unity. As such it's eigenvalues $\lambda_{1,2}$ satisfy the following expression
\begin{equation}
   det(\mathbf{T})=\lambda_1 \lambda_2 =1 \label{det}.
\end{equation}
Then, since here $\lambda_1=\mathrm{e}^{-\mathrm{i}2q \alpha}$, for real values of $q\alpha$ the second eigenvalue is the complex conjugate. Finally using the fact that the trace of the matrix is given by the sum of its eigenvalues we directly see that 
 \begin{equation}
   tr(\mathbf{T})=2\cos{(2q\alpha)}. \label{trace}
\end{equation}
From the last expression we obtain the dispersion relation of the periodic acoustic waveguide (see Eq.\,(\ref{transfer matrix})), which relates the non-dimensional Bloch wavenumber $q\alpha$ with the physical frequency using the transfer matrix method.
Dissipative effects in the TMM are included upon using a complex 
form for the wavenumber and the characteristic impedance of each waveguide segment respectively, namely:  
%
\begin{align}
        &k_j=\frac{\omega}{c_0}\Big(1 + \frac{1-\mathrm{i}}{s}\big(1 + (\gamma -1)\big)/\sqrt{Pr}\Big),\label{wavenumberlosses} \\
        &Z_j=\frac{\rho_0 c_0}{S_j}\Big(1 + \frac{1-\mathrm{i}}{s}\big(1 - (\gamma- 1)\big)/\sqrt{Pr}\Big), \label{impedancelosses}
    \end{align}
where $\omega$ is the frequency, $c_0$ is the speed of sound in air, $\gamma$ is the specific heat ratio, $\Pr$ is the Prandtl number, $s=\sqrt{\omega \rho h^{2}_{j}/\eta}$ and $\eta$ is the shear viscosity.

\begin{figure}[tbp]
\begin{center}
   \includegraphics[width=0.4\textwidth]{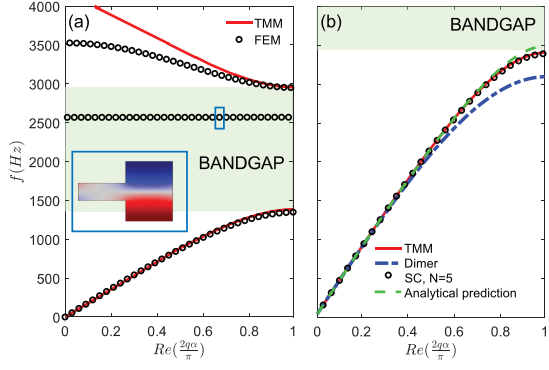}
  \caption{(a) The dispersion relation of the periodic waveguide as obtained with the TMM (red) and the 2D Helmholtz (black) with FEM. (b) The dispersion relation of the periodic waveguide as obtained with the TMM (red), the diatomic TL (blue), the supercell TL (black) and the effective PDE dispersion (green).
  \label{dispersion2}}
  \end{center}
\end{figure}

In order to verify the 1D approximation, we compute the dispersion of the periodic waveguide by solving the 
fully 2D Helmholtz equation using FEM, and we compare it with the expression\,(\ref{transfer matrix}) (in the lossless case) derived through the TMM.
The results are shown in Fig.\,\ref{dispersion2} (a), where the (black) dots correspond to the FEM  while the (red) lines  correspond to the TMM. As is observed, 
Eq.\,(\ref{transfer matrix}) captures very well the the first branch of the dispersion relation obtained by the simulations,
The second branch, starting at $3K$\,Hz is not captured well by the analytical model, since the 1D approximation at the frequencies starts to fail. However, for the purposes of this work we focus only in the frequency range below the first band gap, justifying a 1D description to be used later.
Note that, the simulations show the existence of a quasi-flat band around $2.5K$\,Hz which is due to the transverse resonance of the large segment as illustrated in the inset of Fig.\,\ref{dispersion2} (a).

\section{Dispersion coefficients of the effective PDE}\label{app2}
Our goal 
is to determine these three parameters so that to effectively 
describe 
the properties of the system under study. Employing the usual ansatz for plane waves in the infinitely small-amplitude regime, $p \propto \exp [\mathrm{i}(q\alpha x-\omega t)]$, we derive from Eq.\,(\ref{nonlinear homogenization}) 
the following dispersion relation: 
\begin{equation}
    \omega=\pm\sqrt{\frac{\tilde{c}^2(q\alpha)^2-\beta_x (q\alpha)^4}{1+\beta_m(q\alpha)^2}}.\label{eq6}
\end{equation}
In the limit of long waves, with $q\alpha\ll 1$, the dispersion relation can be approximated as: 
%
\begin{align}
&\omega \tilde{\alpha}=\tilde{c}_0\bigg(q \alpha-\frac{\beta_{m}+\beta_{x}}{2}(q \alpha)^{3}\nonumber\\
&+\frac{\left(\beta_{m}+\beta_{x}\right)\left(3 \beta_{m}-\beta_{x}\right)}{8}(q \alpha)^{5}+O\left((q \alpha)^{7}\right)\bigg), \label{homogenization dispersion}
\end{align}
where we have kept terms up to the order $O\left((q \alpha)^{7}\right)$.
Now, we use the analytical expression of the dispersion relation\,(\ref{transfer matrix}) in order to find the parameters of the effective Eq.\,(\ref{homogenization dispersion}).
To do so, we find  
the Taylor expansion of the right hand side of Eq.\,(\ref{transfer matrix}) in $k\alpha$ and,   
keeping up to order $O\left((k \alpha)^{4}\right)$, 
we obtain a fourth order polynomial equation for $k\alpha$. Solving the latter, we find $k\alpha$ as a function of $\cos(2q\alpha)$. 
To this end, we expand the cosine function, keeping up to order $O\left((q \alpha)^{6}\right)$ and arrive at:
\begin{equation}
\omega \tilde{\alpha}=\tilde{c}_0\left(q \alpha-\frac{\delta}{2}(q \alpha)^{3}+3\delta(Q-\delta)(q \alpha)^{5}+O\left((q\alpha)^{7}\right)\right), 
\label{Bloch dispersion}
\end{equation}
where 
\begin{eqnarray}
\tilde{c}_0&=&
2\frac{ \sqrt{{Z}_1 {Z}_2}}{{Z}_1+{Z}_2}c_0,
\quad \delta=
\frac{({Z}_1-{Z}_2)^{2}}
{6({Z}_1+{Z}_2)^{2}}, \quad Q=\frac{Q_1}{Q_2},
\nonumber\\
Q_1&=&27 Z_1^{4}-12 Z_1^{3} Z_2-158 Z_1^{2} Z_2^{2}-12 Z_1 
Z_2^{3}+27 Z_2^{4},
\nonumber \\ 
\quad Q_2&=&180(Z_1-Z_2)^{2}(Z_1+Z_2)^{2}.
%
\end{eqnarray}
By direct comparison of Eq.\,(\ref{Bloch dispersion}) with  Eq.\,(\ref{homogenization dispersion}), first we identify at order $qa$ the effective speed of a sound $\tilde{c}_0$. 
Second, to determine 
the rest two coefficients, $\beta_x$ and $\beta_m$, 
we use the physical constraint that at the end of the Brillouin zone the group velocity 
vanishes \cite{wautier}, namely:
\begin{equation}
\left.\frac{\partial \omega}{\partial(q\alpha)}\right|_{q \alpha=\frac{\pi}{2}}=0.
\end{equation}
This gives a first equation connecting 
$\beta_x$ and $\beta_m$ 
[Eq.\,(\ref{homogenization dispersion})].
Then another equation is found from 
Eqs.\,(\ref{homogenization dispersion})  and\,(\ref{Bloch dispersion})  upon equating the fifth order terms~\cite{cornaggia}.
One can then solve these two equations and obtain $\beta_m$ and $\beta_x$. In order to validate our result, in Fig.\,\ref{dispersion2}(b), we compare the effective dispersion relation given by Eq.\,(\ref{eq6}) (dashed (green) line) with the one of TMM (solid (red) line). It is observed that Eq.\,(\ref{eq6}) describes accurately the whole first branch and not only 
the dispersion of the waveguide in the long wavelength limit.

\section{Experimental data}
\textit{Velocity-Amplitude plot.} For each explosion, we obtain four temporal profiles, one corresponding to the source and three to a solitary wave. From these data, we export three different velocities by calculating the time difference of the center of mass, between two consecutive microphones.
For each velocity we associate an amplitude given by the average value of the maximum pressure at the two corresponding consecutive positions. The errorbar is given by assuming an error on the exact position of the microphone holes.

\textit{Width-Amplitude plot.} 
For each measured pulse (at every microphone position) we identify the maximum pressure amplitude and estimate its corresponding halfwidth. The errorbar is given by the deviation of the maximum amplitude due to the inherent noise.


\end{document}